\documentclass[12pt]{article}
\usepackage{amssymb}
\begin{document}



\def\a{\alpha}
\def\b{\beta}
\def\d{\delta}
\def\e{\epsilon}
\def\g{\gamma}
\def\h{\mathfrak{h}}
\def\k{\kappa}
\def\l{\lambda}
\def\o{\omega}
\def\p{\wp}
\def\r{\rho}
\def\t{\tau}
\def\s{\sigma}
\def\z{\zeta}
\def\x{\xi}
\def\V={{{\bf\rm{V}}}}
 \def\A{{\cal{A}}}
 \def\B{{\cal{B}}}
 \def\C{{\cal{C}}}
 \def\D{{\cal{D}}}
\def\G{\Gamma}
\def\K{{\cal{K}}}
\def\O{\Omega}
\def\R{\bar{R}}
\def\T{{\cal{T}}}
\def\L{\Lambda}
\def\f{E_{\tau,\eta}(sl_2)}
\def\E{E_{\tau,\eta}(sl_n)}
\def\Zb{\mathbb{Z}}
\def\Cb{\mathbb{C}}

\def\R{\overline{R}}

\def\beq{\begin{equation}}
\def\eeq{\end{equation}}
\def\bea{\begin{eqnarray}}
\def\eea{\end{eqnarray}}
\def\ba{\begin{array}}
\def\ea{\end{array}}
\def\no{\nonumber}
\def\le{\langle}
\def\re{\rangle}
\def\lt{\left}
\def\rt{\right}

\newtheorem{Theorem}{Theorem}
\newtheorem{Definition}{Definition}
\newtheorem{Proposition}{Proposition}
\newtheorem{Lemma}{Lemma}
\newtheorem{Corollary}{Corollary}
\newcommand{\proof}[1]{{\bf Proof. }
        #1\begin{flushright}$\Box$\end{flushright}}

\renewcommand{\thefootnote}{\fnsymbol{footnote}}
 \setcounter{footnote}{0}

\newfont{\elevenmib}{cmmib10 scaled\magstep1}
\newcommand{\preprint}{
   \begin{flushleft}
   \end{flushleft}\vspace{-1.3cm}
   \begin{flushright}\normalsize
   \end{flushright}}
\newcommand{\Title}[1]{{\baselineskip=26pt
   \begin{center} \Large \bf #1 \\ \ \\ \end{center}}}
\newcommand{\Author}{\begin{center}
   \large \bf

Xiaotian Xu ${}^{a}$,~Junpeng Cao${}^{b,c}$,~Kun Hao ${}^{a}$\footnote{Corresponding author: haoke72@163.com},~
Zhan-Ying Yang${}^{d}$~ and~ Wen-Li Yang${}^{a,e}$\footnote{Corresponding author: wlyang@nwu.edu.cn}

\end{center}}
\newcommand{\Address}{\begin{center}

${}^a$ Institute of Modern Physics, Northwest University,
       Xian 710069, P.R. China\\
${}^b$Beijing National Laboratory for Condensed Matter
           Physics, Institute of Physics, Chinese Academy of Sciences, Beijing
           100190, China\\
${}^c$Collaborative Innovation Center of Quantum Matter, Beijing, China\\
${}^d$Department of Physics, Northwest University, Xi'an 710069, China\\
${}^e$Beijing Center for Mathematics and Information Interdisciplinary Sciences, Beijing, 100048,  China

   \end{center}}
\newcommand{\Accepted}[1]{\begin{center}
   {\large \sf #1}\\ \vspace{1mm}{\small \sf Accepted for Publication}
   \end{center}}

\preprint
\thispagestyle{empty}
\bigskip\bigskip\bigskip
\Title{Small polaron  with generic open boundary conditions: exact solution via the off-diagonal
Bethe ansatz } \Author

\Address
\vspace{1cm}

\begin{abstract}
The small polaron, a one-dimensional lattice model of interacting spinless fermions,   with generic non-diagonal boundary terms is studied
by the off-diagonal Bethe ansatz method. The presence of the Grassmann valued non-diagonal boundary fields gives rise to  a typical
$U(1)$-symmetry-broken fermionic model. The exact spectra of the Hamiltonian and the associated Bethe ansatz equations are derived by constructing
an inhomogeneous $T-Q$ relation.

\vspace{1truecm} \noindent {\it PACS:} 75.10.Jm; 02.30.Ik; 03.65.Fd

\end{abstract}

\newpage

\setcounter{footnote}{0}
\renewcommand{\thefootnote}{\arabic{footnote}}

\baselineskip=20pt

\section{Introduction}

\label{intro} \setcounter{equation}{0}
In this paper we focus on constructing the Bethe ansatz solution of the small polaron with generic non-diagonal boundary terms,
described by the Hamiltonian
\bea
H&=&\sum_{j=1}^{N-1}\frac{1}{\sin\eta}\left\{\cos(\eta)\bar{n}_{j+1} \bar{n}_{j}+
        \cos(\eta) n_{j+1} n_{j}+c^{+}_{j}c_{j+1}+c^{+}_{j+1}c_{j}\right\}\no\\
 &&+\frac{1}{2}\cot(\psi_{-})[\bar{n}_{1}-n_{1}]+[\kappa_{+}\bar{n}_{N}-
        \kappa_{-}n_{N}]+\csc(\psi_{-})\left[\alpha_{-}c_{1}+\beta_{-}c^{+}_{1}\right]\no\\
 &&+\csc(\psi_{+})\left[\alpha_{+}c_{N}+\beta_{+}c^{+}_{N}\right],\label{Hamiltonian}
\eea where $c_{j}^{+}$ and $c_{j}$ are the creation and annihilation operators of spinless fermions at site $j$ (which obey anticommutation relations $\{c_{j}^{+},c_{k}\}=\delta_{jk}$), respectively; the operators of particle numbers are $n_{j}=c_{j}^{+}c_{j}$ and $\bar{n}_{j}=1-n_{j}$; the
parameters $\psi_{\pm}$, $\a_{\pm}$ and $\b_{\pm}$ are the boundary parameters related to boundary interactions;  $\eta$ is the bulk coupling parameter. The boundary coupling $\kappa_{\pm}$ is given by  $\frac{1}{2}\csc\psi_{+}\csc\eta\sin(\eta\pm\psi_{+})$ respectively.
The  model (\ref{Hamiltonian}) is a typical  spinless fermion model with boundary terms in condensed matter physics. It provides an effective description of the motion of an additional electron in a polar crystal \cite{Fed78,Mak84}. In one spatial dimension, the model is integrable for both periodic and open boundary conditions by reconstructing it
within the framework of quantum inverse scattering method (QISM) \cite{Skl80, Pu86, Skl88, Bra98}.

In the past few decades, the integrability and the excitation spectrums problem have been studied extensively.
For the small polaron model with periodic and purely diagonal boundary conditions, which can be mapped onto the XXZ quantum spin chain through the Jordan-Wigner transformation, the energy spectrum problem of the model was solved by the Algebra Bethe Ansatz method in \cite{Guan99,Fan97,Ume99}.  A remarkable result was given by Yukiko Umeno \cite{Ume99} who constructed the fermionic R-operator and solved the spectrum problem via the Algebra Bethe Ansatz method. The generic integrable boundary conditions were obtained \cite{Zhou96} by solving the graded reflection  equation \cite{Che84}. Subsequently, the Lax pair formulation of the generic integrable boundary conditions was presented in \cite{Guan98}.  Since then, there have been numerous efforts  to work out the exact solutions of the model. In 2013 the authors in \cite{Gabi13, Kar13} figured out the Bethe ansatz solution of the model with non-diagonal boundary terms based on a deformation of the diagonal case  and commented on the eigenstate of the model which envolves into the Fock vacuum when the off-diagonal boundary terms were ignored. The result is also closely related to that of algebra Bethe ansatz method.
However, there still exists a main obstacle for applying the conventional algebra Bethe Ansatz method to get the exact solution of the model with generic
off-diagonal boundary conditions. The difficulty is mainly due to the fact the Hamiltonian (\ref{Hamiltonian}) includes Grassmann valued non-diagonal boundary fields
(or couplings) such as the terms associated with the parameters $\a_{\pm}$ and $\b_{\pm}$ which breaks the bulk $U(1)$-symmetry of the model.
The breaking of the $U(1)$-symmetry  leads to the obvious reference state (all-spin-up or all-spin-down state) is no longer the reference state in the usual algebraic
Bethe ansatz \cite{Kor93}.

Very recently, a systematic method for  approaching the exact solutions of generic integrable models either with $U$(1) symmetry or not,
i.e., the off-diagnonal Bethe ansatz (ODBA) method \cite{Wan15} was proposed in \cite{Cao1,Cao2,Cao3,Cao4}. With the ODBA method, some long-standing models \cite{Cao3,Li14,Zha14,Hao14, Cao15} without $U$(1) symmetry were then solved. In this paper we study the small polaron model with the generic integrable boundary condition specified by the $K$-matrices within Grassmann numbers via the ODBA  method.

The paper is organized as follows. In Section 2, we begin with a concise view of the integrability of the fermion model with
the open boundary condition  within the framework of the graded QISM. Some
basic ingredients and algorithm of the transfer matrix are also introduced.
In Section 3  we show that the Hamiltonian of the model can be  rewritten in terms of the corresponding transfer matrix.
In Section 4, after deriving the operator product identities of the transfer matrix at some special  points of the spectrum parameter and its asymptotic behaviors,
we express the eigenvalue  of the transfer matrix in terms of  an inhomogeneous  $T-Q$ relation  and  derive the associated Bethe ansatz equations. Finally, we summarize our results and give some discussions .
\section{Transfer matrix}
\setcounter{equation}{0}

Let  V be a two-dimensional $\mathbb{Z}_{2}$-graded  vector space (or super space) \cite{Fra96} with an orthnormal basis $\{|i\rangle|i=1,2\}$. The grading of the
basis vectors is $[|1\rangle]=0$, $[|2\rangle]=1$. The R-matrix of the small polaron model is given by \cite{Kar13}
\bea
R(u)=\frac{1}{\sin\eta}\left(
\begin{array}{cccc}
  \sin(u+\eta) & 0 & 0 & 0 \\
  0 & \sin u & \sin \eta & 0 \\
  0 & \sin \eta & \sin u & 0 \\
  0 & 0 & 0 & -\sin(u+\eta)
\end{array}\right),
\eea
acting on the tensor product $V\otimes V$ of two superspace.
Here $u$ is the spectral parameter and  $\eta$ is the crossing parameter related to the bulk coupling (\ref{Hamiltonian}). The $R$-matrix
$R(u)$ satisfies  the graded quantum Yang-Baxter equation (g-QYBE) \cite{Kul86}
\bea
R_{12}(u-v)R_{13}(u)R_{23}(v)=R_{23}(v)R_{13}(u)R_{12}(u-v),\label{ybe}
\eea
and enjoys the properties:
\bea &&\hspace{-1.5cm}\mbox{Initial
condition}:\,R_{12}(0)= P_{12},\label{Int-R}\\
&&\hspace{-1.5cm}\mbox{Unitarity relation}: R_{12}(u)R_{21}(-u)=\xi(u),\quad \xi(u)=-\frac{\sin(u-\eta)}{\sin\eta}\frac{\sin(u+\eta)}{\sin\eta}, \\
&&\hspace{-1.5cm}\mbox{P-symmetry}:
R_{21}(u)=P_{12}R_{12}(u)P_{12}=R_{12}(u),\\
&&\hspace{-1.5cm}\mbox{T-symmetry}:
R_{12}^{st_{1},\,st_{2}}(u)=R_{12}^{ist_{1},\,ist_{2}}(u)=R_{21}(u),\\
&&\hspace{-1.5cm}\mbox{Crossing relation}:
R_{21}^{st_{2}}(-u-2\eta)R_{21}^{st_{1}}(u)=\xi(u+\eta),\\
&&\hspace{-1.5cm}\mbox{Antisymmetry}:
R_{12}(-\eta)=-2P^{(-)},\label{anti}\\
&&\hspace{-1.5cm}\mbox{Periodicity}:
R_{12}(u+\pi)=-\sigma_{1}^{z}R_{12}(u)\sigma_{1}^{z}=-\sigma_{2}^{z}R_{12}(u)\sigma_{2}^{z}.
\label{peri}\eea
In the above equations, $st_{j}$ and $ist_{j}$ are the partial super transposition and its inverse,
$P_{ij}$ is the graded permutation operator and $P^{(-)}$ is a projector with rank one,
\bea
P^{(-)}=\frac{1}{2}\left(
\begin{array}{cccc}
  0 & 0 & 0 & 0 \\
  0 & 1 & -1 & 0 \\
  0 & -1 & 1 & 0 \\
  0 & 0 & 0 & 0
\end{array}\right).
\eea
Here
and below we adopt the standard notations: for any matrix $A\in {\rm
End}({ V})$, $A_j$ is an embedding operator in the tensor
space ${ V}\otimes { V}\otimes\cdots$, which acts as $A$ on the
$j$-th space and as identity on the other factor spaces; $R_{ij}(u)$
is an embedding operator of $R$-matrix in the tensor space, which
acts as identity on the factor spaces except for the $i$-th and
$j$-th ones. Since we discuss a fermionic lattice model, all embeddings
are to be understood into a super tensor product structure. It is remarked that
the super tensor product is graded according to the rule
\bea
(A\otimes B)^{ik}_{~~jl}=(-1)^{([|i\rangle]+[|j\rangle])[|k\rangle]}A^i_{~j}B^k_{~l},
\eea
where the parity $[|i\rangle]$ is equal to zero (one) for bosonic (fermionic) indices.
(For the details about
the algorithm of super tensor product we refer the reader to \cite{Fra96, Gabi13}.)

We introduce two  monodromy matrices $T_{0}(u)$ and $\hat{T}_{0}(u)$, which can be considered as 2$\times$2 matrices on the auxiliary space with elements being operators acting on $V^{{\otimes}^{N}}$,
\bea &&\hspace{-1.5cm}
T_{0}(u)=R_{0N}(u-\theta_{N})R_{0N-1}(u-\theta_{N-1})\cdots R_{01}(u-\theta_{1})\label{tu},\\
&&\hspace{-1.5cm}\hat{T}_{0}(u)=R_{01}(u+\theta_{1})R_{02}(u+\theta_{2})\cdots R_{0N}(u+\theta_{N}).
\eea
Here $\{\theta_{j}|j=1,2,\cdots,N\}$ are arbitrary free complex parameters which are usually called the inhomogeneous parameters.

The framework of QISM for integrable systems with open boundary conditions  in a way that makes it applicable to super spin chains. Following \cite{Skl88,Bra98}, for a given R-matrix, we introduce a pair of K-matrices $K^{-}(u)$ and $K^{+}(u)$. The former satisfies the graded reflection equation
\bea
&&R_{12}(u-v)K_{1}^{-}(u)R_{21}(u+v)K_{2}^{-}(v)\no\\
 &&~~~~~~=
 K_{2}^{-}(v)R_{12}(u+v)K_{1}^{-}(u)R_{21}(u-v)\label{ref},
\eea
and the latter satisfies the dual graded reflection equation
\bea
&&R_{12}(v-u)K_{1}^{+}(u)\tilde{\tilde{R}}_{21}(-u-v)^{ist_{1},\,st_{2}}K_{2}^{+}(v)\no\\
 &&~~~~~~~~~~~=
K_{2}^{+}(v)\tilde{R}_{12}(-u-v)^{ist_{1},\,
st_{2}}K_{1}^{+}(u)R_{21}(v-u)\label{dul ref},
\eea
 whereas the new matrices $\tilde{\tilde{R}}$ and $\tilde{R}$ are related to the R-matrix via
\bea &&\hspace{-1.5cm}
\tilde{\tilde{R}}{_{21}(u)^{ist_{1},\,st_{2}}=([\{{R_{21}^{-1}(u)}\}}^{ist_{2}}]^{-1})^{st_{2}},\\[6pt]
&&\hspace{-1.5cm}\tilde{R}_{12}(u)^{ist_{1},\,st_{2}}=([\{R_{12}^{-1}(u)\}^{st_{1}}]^{-1})^{ist_{1}}.
\eea

For open super spin chains, rather than the standard monodromy matrix $T_{0}(u)$ (\ref{tu}), we need to consider the double-row monodromy matrix $\mathbb{T}_{0}(u)$
\bea
\mathbb{T}_{0}(u)=T_{0}(u)K_{0}^{-}(u)\hat{T}_{0}(u).
\eea
Then the double-row transfer matrix $t(u)$ of the system is given by
\bea
t(u)=str_{0}\{K_{0}^{+}(u)\mathbb{T}_{0}(u)\}, \label{com}
\eea
where $str\{\cdot\}$ denotes the super trace of a matrix, which
is defined by
\bea
str\{A\}\equiv\sum_i(-1)^{[i]}A^i_i.
\eea
The graded QYBE (\ref{ybe}) and REs (\ref{ref}) and (\ref{dul ref}) lead to the fact that the transfer matrices  give rise to a family of commuting operators \cite{Bra98} with different spectral parameters:
\bea
[t(u),\,t(v)]=0.
\eea
Then $t(u)$ serves as the generating function of the conserved quantities, which ensures the integrability of the system.

\section{Small polaron with open boundaries}
\setcounter{equation}{0}

In this paper, we consider the K-matrices $K^{-}(u)$ and $K^{+}(u)$ which satisfy the graded REs \cite{Bra98,Skl88} and possess the following generic expressions (see also \cite{Zhou96,Zhou97,Guan98})
\bea
K^{-}(u)=\omega_{-}\left(
\begin{array}{cc}
  \sin(u+\psi_{-}) & \alpha_{-}\sin(2u) \\
  \beta_{-}\sin(2u)  &  -\sin(u-\psi_{-})
\end{array}\right)\label{k-},
\eea
\bea
K^{+}(u)=\omega_{+}\left(
\begin{array}{cc}
  \sin(u+\eta+\psi_{+}) & \alpha_{+}\sin(2[u+\eta]) \\
  \beta_{+}\sin(2[u+\eta])  &  \sin(u+\eta-\psi_{+})
\end{array}\right)\label{k+},
\eea
with normalizations $\omega_{\pm}$ defined by
$\omega_{-}(\eta)\equiv\frac{1}{\sin(\psi_{-})} $ and $\omega_{+}(\eta)\equiv\frac{1}{2\cos(\eta)\sin(\psi_{+})}
$. Here $\psi_{\pm}$, $\alpha_{\pm}$, $\beta_{\pm}$ are all Grassmann numbers which are related to boundary fields. The parameters $\psi_{\pm}$ are arbitrary commuting  even Grassmann numbers  but the invertibility  requires them to have a non-vanishing complex part, the remaining non-diagonal boundary parameters $\alpha_{\pm}$ and $\beta_{\pm}$ are anticommuting odd Grassmann numbers, namely, 
\bea
[\psi_{+},\psi_{-}]=0=\{\alpha_{\pm},\alpha_{\pm}\}=\{\alpha_{\pm},\beta_{\pm}\}=\{\beta_{\pm},\beta_{\pm}\}.
\eea
In addition, the odd Grassmann numbers are subject to the condition $\alpha_{\pm}\beta_{\pm}=0$ due to  the graded REs (\ref{ref}) and (\ref{dul ref}).

Based on the graded QISM, the Hamiltonian (\ref{Hamiltonian}) of the small polaron model
with generic off-diagonal boundary terms can be rewritten in terms of the transfer matrix (\ref{com}) as:
\bea
H&=&\frac{1}{2}\frac{\partial{t(u)}}{\partial{u}}|_{{u=0},\{\theta_{j}=0\}}+\frac{1}{2}\tan\eta\no\\
&=&\frac{1}{2}str_{0}\{K^{+^{'}}_{0}(0)\}+\sum_{j=1}^{N-1}R^{'}_{j,j+1}(0)P_{j,j+1}+str_{0}\{K_{0}^{+}(0)
P_{N0}R^{'}_{0N}(0)\}\no\\
&&+K_{1}^{-^{'}}(0)+\frac{1}{2}\tan\eta.\label{ham}\eea
The purpose of this paper is to construct the spectra of  the  Hamiltonian  and derive the corresponding Bethe ansatz equations.
\setcounter{equation}{0}
\section{Eigenvalues and the Bethe ansatz equations}

\subsection{Functional relations}
Following the similar method developed in \cite{Cao4}, we derive that the products of the transfer matrix (\ref{com}) of the super spin chain with the generic open boundaries described by the K-matrices in (\ref{k-}) and (\ref{k+}), at the points $\theta_j$ and $\theta_j-\eta$,  satisfies the relations
\bea
t(\theta_{j})t(\theta_{j}-\eta)=-\frac{\Delta_{q}(\theta_{j})}{\xi(2\theta_{j})}, \quad j=1,\ldots,N.\label{ope}
\eea
For generic $\{\theta_{j}\}$, the quantum determinant operator $\Delta_{q}(\theta_{j})$ is proportional to the identity operator, namely,
\bea
\Delta_{q}(u)=\delta(u)\times {\rm id},
\eea
where the function $\delta(u)$  is given by
\bea
\delta(u)&=&\frac{\omega_+^2\omega_-^2}{\sin^2{\eta}}\sin(u+\psi_+)\sin(u-\psi_+)\sin(u+\psi_-)\sin(u-\psi_-)\sin(2u+2\eta)\sin(2u-2\eta)\no\\[6pt]
&&\quad\times\prod_{l=1}^N\frac{\sin(u-\theta_l-\eta)\sin(u-\theta_l+\eta)}{\sin^2{\eta}}\frac{\sin(u+\theta_l-\eta)\sin(u+\theta_l+\eta)}{\sin^2{\eta}}
\label{u}.\eea
Furthermore, we have checked that the transfer matrix $t(u)$ of the small polaron model with the generic boundary conditions enjoys the crossing property
\bea
t(-u-\eta)=t(u)\label{cro}.
\eea
The quasi-periodicity of the R-matrix (\ref{peri}) and K-matrices
\bea
R_{12}(u+\pi)=-\sigma_1^z R_{12}(u)\sigma_1^z=-\sigma_2^z R_{12}(u)\sigma_2^z,\quad K^{\pm}(u+\pi)=-\sigma^z K^{\pm}(u)\sigma^z\label{rk},
\eea
and the special points values at $u=0, \frac{\pi}{2}$ of the K-matrix give rise to several properties of the associated  transfer matrix, namely,
\bea
t(u+\pi)&=&t(u)\label{t},\\
t(0)&=&\prod_{l=1}^N\frac{\sin(\eta-\theta_l)\sin(\eta+\theta_l)}{\sin^2 \eta}\times {\rm id},\\
t(\frac{\pi}{2})&=&\cot\psi_{-}\cot\psi_{+}\prod_{l=1}^N\frac{\sin(\frac{\pi}{2}-\theta_l+\eta)\sin(\frac{\pi}{2}+\theta_l+\eta)}{\sin^2\eta}\times{\rm id},\\
\lim_{iu\rightarrow \pm\infty}t(u)&=&\omega_+\omega_-\frac{1}{(2i)^{2N+2}}(\alpha_+\beta_{-}-\beta_+\alpha_-)\frac{1}{\sin^{2N}\eta}e^{\pm\{i(2N+4)u+i(N+2)\eta\}} \times U^z. \label{asy}
\eea
Here the operator $U^z$ is given by
\bea
U^z=\prod_{j=1}^N\sigma_{j}^z,\quad (U^z)^2={\rm id}, \label{U}
\eea which commutes with the transfer matrix. The relation (\ref{U}) allows us to decompose the whole Hilbert space $\cal{H}$ into two subspaces, i.e., $\cal{H}= \cal{H}^+\oplus\cal{H}^-$ according to the action of the operator $U^z$: $U^z\,\cal{H}^{\pm}=\pm \cal{H}^{\pm}$. The commutativity of the transfer matrix and the
operator $U^z$, i.e., $[t(u),\,U^z]=0$, implies that each of the subspace is invariant under $t(u)$. Hence the whole set of eigenvalues of the transfer matrix
can be decompose into two series, denoted by $\L_{\pm}(u)$ respectively. The eigenstates corresponding to $\L_{+}(u)$ (resp.  $\L_{-}(u)$ ) belong to the subspace $\cal{H}^+$ (resp. $\cal{H}^-$). The operator product identities (\ref{ope}) of the transfer matrix and the commutativity of the transfer matrix with different
spectrum $u$ enable us to derive the following relations of the associated eigenvalues $\Lambda_{\pm}(u)$ respectively,
\bea
\Lambda_{\pm}(\theta_j)\Lambda_{\pm}(\theta_j-\eta)=\frac{\delta(\theta_j)\sin\eta \sin\eta}{\sin(2\theta_j+\eta)\sin(2\theta_j-\eta)}\label{lam},
~~~j=1,\cdots, N,
\eea
with the function $\delta(u)$  given in (\ref{u}).

The properties of the transfer matrix $t(u)$ given by (\ref{cro})-(\ref{asy}), imply that the corresponding eigenvalue functions $\Lambda_{\pm}(u)$ satisfy the relations:
\bea
\Lambda_{\pm}(-u-\eta)&=&\Lambda_{\pm}(u),\quad \Lambda_{\pm}(u+\pi)=\Lambda_{\pm}(u),\label{cr}\\[6pt]
\Lambda_{\pm}(0)&=&\prod_{l=1}^N\frac{\sin(\eta-\theta_l)\sin(\eta+\theta_l)}{\sin^2 \eta},\\[6pt]
\Lambda_{\pm}(\frac{\pi}{2})&=&\cot\psi_{-}\cot\psi_{+}\prod_{l=1}^N\frac{\sin(\frac{\pi}{2}-\theta_l
+\eta)\sin(\frac{\pi}{2}+\theta_l+\eta)}{\sin^2\eta},\\[6pt]
\lim_{iu\rightarrow \pm\infty}\Lambda_{\pm}(u)&=&\pm\quad\omega_+\omega_-\frac{1}{(2i)^{2N+2}}
(\alpha_+\beta_{-}-\beta_+\alpha_-)\frac{1}{\sin^{2N}\eta}e^{\pm \{i(2N+4)u+i(N+2)\eta\}}.\label{lasy}
\eea
Obviously, $\Lambda_{\pm}(u)$ are a degree $2N+4$ trigonometric polynomial of $u$, along with the crossing symmetry and the periodicity mentioned in (\ref{cr}), these factors lead to that only $N+3$ unknown coefficients need to be determined by $N+3$ special points values of the associated function $\Lambda_{\pm}(u)$.
 Therefore, the two functions $\Lambda_{\pm}(u)$ can be completely determined by the above functional relations (\ref{lam})-(\ref{lasy}).
\subsection{Eigenvalues of the transfer matrix }
Following the method in \cite{Cao1,Cao2,Cao3,Cao4} and with the helps of the functional relations (\ref{lam})-(\ref{lasy}), we can express the eigenvalue
$\L_{\pm}(u)$ of the transfer matrix of the small polaron model with the boundary terms specified by the generic non-diagonal K-matrices given by
 (\ref{k-}) and (\ref{k+}) in terms of an inhomogeneous $T-Q$ relation \cite{Cao2} respectively,
\bea
\Lambda_{\pm}(u)&=&a(u)\frac{Q^{(\pm)}(u-\eta)}{Q^{(\pm)}(u)}+d(u)\frac{Q^{(\pm)}(u+\eta)}{Q^{(\pm)}(u)}
\no\\[6pt]
&&\quad\pm\frac{\bar{c}\sin(2u)\sin(2u+2\eta)}{Q^{(\pm)}(u)}\bar{A}(u)\bar{A}(-u-\eta)
\label{eig},\eea
where the Q-functions are parameterized by $\{\mu^{(\pm)}_{j}\mid j=1,\cdots,N\}$ respectively
\bea
Q^{(\pm)}(u)&=&\prod_{j=1}^{N}\frac{\sin(u-\mu^{(\pm)}_j)}{\sin\eta}\frac{\sin(u+\mu^{(\pm)}_j+\eta)}{\sin\eta}
=Q^{(\pm)}(-u-\eta).\label{Q-function-1-1}
\eea
The other functions $a(u)$, $d(u)$, $\bar{A}(u)$ and the constant $\bar{c}$ are given by
\bea
\bar{A}(u)&=&\prod_{l=1}^N\frac{\sin(u-\theta_l+\eta)\sin(u+\theta_l+\eta)}{\sin^2 \eta},\\
 a(u)&=&\omega_+\omega_-\sin(u-\psi_+)\sin(u-\psi_-)\frac{\sin(2u+2\eta)}{\sin(2u+\eta)}\bar{A}(u),\\
 d(u)&=&\omega_+\omega_-\sin(u+\eta+\psi_+)\sin(u+\eta+\psi_-)\frac{\sin(2u)}{\sin(2u+\eta)}
 \bar{A}(-u-\eta)\no\\
 &=&a(-u-\eta),\\
 \bar{c}&=&\omega_+\omega_-(\alpha_+\beta_{-}-\beta_+\alpha_-).
\eea
Since that $\Lambda_{\pm}(u)$ both are polynomials, the residues of $\Lambda_{\pm}(u)$ at the apparent poles $u=\mu^{(\pm)}_j$ and $u=-\mu^{(\pm)}_j-\eta$, $j=1,\cdots,N$ must vanish, which gives rise to  the associated BAEs
\bea
&&\hspace{-1.2truecm}a(\mu^{(\pm)}_j)Q(\mu^{(\pm)}_j-\eta)+d(\mu^{(\pm)}_j)
Q(\mu^{(\pm)}_j+\eta)\no\\[6pt]
&&\hspace{-1.2truecm}\quad\quad\pm\bar{c}\sin2\mu^{(\pm)}_j\sin(2\mu^{(\pm)}_j+2\eta)\bar{A}(\mu^{(\pm)}_j)\bar{A}(-\mu^{(\pm)}_j-\eta)=0,\quad j=1,\ldots,N.\label{bae}
\eea
It is easy to check that the $T-Q$ relation (\ref{eig}) satisfies the relations (\ref{lam})-(\ref{lasy}) respectively  under the condition of $N$ parameters $\{\mu_{j}\mid j=1,\cdots,N\}$ satisfying the BAEs (\ref{bae}).

We remark that the roots $\{\mu^{(\pm)}_j|j=1,\cdots,N\}$ to the BAEs (\ref{bae}) are Grassmann number valued, which  implies that the corresponding $Q$-functions in (\ref{Q-function-1-1}) can be expressed as
\bea
Q^{(\pm)}(u)=Q_0^{(\pm)}(u)+g\,Q_1^{(\pm)}(u),\quad g=\alpha_+\beta_{-}-\beta_+\alpha_-,\, {\rm and}\, g^2=0,\label{Q-1}
\eea
where
\bea
Q_0^{(\pm)}(u)&=&\prod_{j=1}^{N}\frac{\sin(u-\l^{(0,\pm)}_j)}{\sin\eta}\frac{\sin(u+\l^{(0,\pm)}_j+\eta)}{\sin\eta},\no\\
Q_1^{(\pm)}(u)&=&\l^{(1,\pm)}_N\prod_{j=1}^{N-1}\frac{\sin(u-\l^{(1,\pm)}_j)}{\sin\eta}\frac{\sin(u+\l^{(1,\pm)}_j+\eta)}{\sin\eta}.\label{Q-2}
\eea
The $2N$  parameters $\{\l^{(i,\pm)}_j|i=0,1;\,j=1\cdots,N\}$ are c-number valued.
Substituting the relations (\ref{Q-1})-(\ref{Q-2}) into the BAEs (\ref{bae}), one may get the associated $2N$ BAEs. The resulting BAEs completely determine the $2N$ c-number valued parameters $\{\l^{(i,\pm)}_j|i=0,1;\,j=1\cdots,N\}$, which resembles those in \cite{Gabi13,Kar13}.

In the homogeneous limit $\theta_j\rightarrow 0$, the above BAEs become
\bea
\hspace{-1.2truecm}&&\lt(\frac{\sin(\mu^{(\pm)}_j+\eta)}{\sin\mu^{(\pm)}_j}\rt)^{2N}\frac{\sin(\mu^{(\pm)}_j-\psi_+)
\sin(\mu^{(\pm)}_j-\psi_-)\sin(2\mu^{(\pm)}_j+2\eta)}{\sin(\mu^{(\pm)}_j+\eta+\psi_+)\sin(\mu^{(\pm)}_j+\eta+\psi_-)\sin(2\mu^{(\pm)}_j)}=
-\frac{Q(\mu^{(\pm)}_j+\eta)}{Q(\mu^{(\pm)}_j-\eta)}\no\\[6pt]
\hspace{-1.2truecm}&&\quad\quad\mp
\frac{(\alpha_+\beta_-\hspace{-0.04truecm}-\hspace{-0.04truecm}\beta_+\alpha_-)
\sin(2\mu^{(\pm)}_j\hspace{-0.04truecm}+\hspace{-0.04truecm}\eta)
\sin(2\mu^{(\pm)}_j\hspace{-0.04truecm}+\hspace{-0.04truecm}2\eta)
\sin^{2N}(\mu^{(\pm)}_j\hspace{-0.04truecm}+\hspace{-0.04truecm}\eta)}
{\sin(\mu^{(\pm)}_j\hspace{-0.04truecm}+\hspace{-0.04truecm}\eta\hspace{-0.04truecm}+\hspace{-0.04truecm}\psi_+)
\sin(\mu^{(\pm)}_j\hspace{-0.04truecm}+\hspace{-0.04truecm}\eta\hspace{-0.04truecm}+\hspace{-0.04truecm}\psi_-)
\sin^{2N}\eta Q(\mu^{(\pm)}_j\hspace{-0.04truecm}-\hspace{-0.04truecm}\eta)},\,\,j=1,\ldots,N.\label{BAE-h}
\eea
Then two series eigenvalues of the Hamiltonian (\ref{Hamiltonian}) can be expressed in terms of the Bethe roots as follows
\bea
E_{\pm}&=&\frac{1}{2}\frac{\partial\Lambda_\pm(u)}{\partial u}\mid_{{u=0},\{\theta_j=0\}}+\frac{1}{2}\tan\eta\no\\
&=&-\frac{1}{2}\cot{\psi_+}-\frac{1}{2}\cot{\psi_-}-\frac{1}{\sin2\eta}+N\cot\eta+
+\frac{1}{2}\tan\eta\no\\
&&+\sum_{j=1}^{N}
\frac{\sin^{2}\eta}{\sin\mu^{(\pm)}_j\sin(\mu^{(\pm)}_j+\eta)}
\label{e},
\eea
where the parameters $\{\mu_j^{(\pm)}|j=1,\ldots,N\}$ satisfy the associated BAEs (\ref{BAE-h}).

\section{Conclusions}
The small polaron model with off-diagonal boundary described by the K-matrices (\ref{k-}) and (\ref{k+}), which can be regarded as a graded version of the general open XXZ spin chain, has been studied by the off-diagonal Bethe ansatz method proposed in \cite{Cao2,Cao1,Cao3,Cao4}. Based on some properties of the R-matrix and K-matrices, we obtain the operator identities (\ref{ope}) of the transfer matrix and then  construct the corresponding inhomogeneous $T-Q$ relation for its eigenvalues (\ref{eig}) and the corresponding BAEs (\ref{bae}). Moreover, the exact spectra of the Hamiltonian is given in (\ref{e}). When the nondiagonal boundary parameters satisfy the constraint $\alpha_{\pm}=\beta_{\pm}=0$, the resulting $T-Q$ relation is reduced to the conventional one which is the solution of the model with diagonal boundaries.

A possible extension of the present work is to consider the mulit-component Bose-Fermi mixtures with off-diagonal boundary conditions with the help of the fusion method \cite{Mez92}. Meanwhile,
according to the spin-$s$  XXZ Heisenberg chain with generic
non-diagonal boundaries solved in \cite{Luc07}, the construction and
the solution of graded higher spin chain may be obtained by similar
method.

\section*{Acknowledgments}

This work has been  financial supported by NSFC under Grant Nos. 11375141, 11374334, 11434013 and 11425522, the 973 project under Grant
No. 2011CB921700, BCMIIS and the Strategic Priority Research Program
of CAS. One of the authors X. Xu was also partially supported by the NWU graduate student
innovation fund No. YZZ14102.

\end{document}